\documentclass[
10pt,
aps,
preprint,
superscriptaddress,
floatfix,
english,
showkeys,
notitlepage,
nofootinbib,
showpacs,
byrevtex
]{revtex4-1}
\usepackage{graphicx}
\usepackage{amsmath,amssymb}
\usepackage{pdfsync}
\usepackage{hyperref}

\usepackage[math]{iwona}
\newcommand{\visco}{\eta_{\scriptscriptstyle \mathrm{V}}}
\newcommand{\resisto}{\eta_{\scriptscriptstyle \mathrm{B}}}
\newcommand{\grad}{\vec{\nabla} \!}
\newcommand{\DELTAE}[1]{\Delta_\beta #1 }
\newcommand{\DERO}[2]{\displaystyle\frac{\partial #1}{\partial #2}}
\newcommand{\DER}[3]{ \displaystyle\frac{\partial^{#3} #1}{\partial #2^{#3}}}
\newcommand{\DELTA}[1]{\left(\DER{#1}{r}{2} + \DER{#1}{z}{2}\right)}
\newcommand{\DERT}[3]{\displaystyle\frac{\mathrm{d}^{#3} #1}{\mathrm{d} #2^{#3}}}
\newcommand{\<}[1]{\! \left( #1 \right)}
\newcommand{\abs}[1]{\left| #1 \right|}
\newcommand{\virg}[1]{`#1'}
\newcommand{\df}[2]{\displaystyle\frac{#1}{#2}}
\newcommand{\SUBEQ}[1]{\begin{subequations} #1 \end{subequations}}
\newcommand{\reff}[1]{(\ref{#1})}
\newcommand{\eref}[1]{Eq.\reff{#1}}
\newcommand{\manualref}[2]{Figure (\hyperref[#1]{#2})}
\newcommand{\Sref}[1]{Section \ref{#1}}

\newcommand{\PRE}{Phys. Rev. E}

\newcommand{\mnras}{Mon. Not. R. Astron. Soc.}
\newcommand{\aap}{Astron. Astrophys.}
\newcommand{\grg}{Gen. Relativ. Gravit.}
\newcommand{\PP}{Phys. Plasma}
\newcommand{\nar}{New Astron. Rev.}
\newcommand{\araa}{Annu. Rev. Astron. Astrophys.}
\newcommand{\SA}{Sov. Astron.}
\newcommand{\epl}{Europhys. Lett.}
\newcommand{\ppcf}{Plasma Phys. Controlled Fusion}

\begin{document}

\title{Non-stationary magnetic microstructures in stellar thin accretion discs}

\author{Giovanni Montani}
\email{giovanni.montani@frascati.enea.it} 
\affiliation{ENEA -- C.R. Frascati, U.T. Fus. (FUSMAG Lab) -- Via Enrico Fermi 45, (00044) Frascati (RM), Italy}
\affiliation{Physics Department, `Sapienza' University of Rome -- Piazzale Aldo Moro 5, (00185) Roma, Italy}

\author{Jacopo Petitta}
\email{petitta.jacopo@gmail.com}
\affiliation{Physics Department, `Sapienza' University of Rome -- Piazzale Aldo Moro 5, (00185) Roma, Italy}

\date{Submitted to PRE on \today}

\begin{abstract}
We examine the morphology of magnetic structures in thin plasma accretion discs, generalizing a stationary ideal MHD model to the time-dependent visco-resistive case.
Our analysis deals with small scale perturbations to a central dipole-like magnetic field, which give rise -- as in the ideal case -- to the periodic modulation of magnetic flux surfaces along the radial direction, corresponding to the formation of a toroidal current channels sequence.
These microstructures suffer an exponential damping in time because of the non-zero resistivity coefficient, allowing us to define a configuration lifetime which mainly depends on the midplane temperature and on the length scale of the structure itself.
By means of this lifetime we show that the microstructures can exist within the inner region of stellar discs in a precise range of temperatures ($R \lesssim 10^9 \, \mathrm{cm}$, $10^4 \, \mathrm{K} \lesssim T \lesssim 10^5 \, \mathrm{K}$), and that their duration is consistent with local transient processes (minutes to hours).
\end{abstract}


\maketitle

\section{Introduction}

A new perspective on the equilibrium morphology of 
thin accretion discs has been introduced 
in \cite{C05,CR06}, where the effect of the plasma 
back-reaction on the magnetic field of the central object is outlined:
it can induce a radially oscillating `crystal profile', meaning that the inner field acquires a well-defined periodic behaviour. This could be described via the formation of small-scale structures in the magnetic surfaces 
(like the one shown in \manualref{fig:crystal}{2}; see also \cite{LM10,MB11}), and it is inferred that such microstructures distort the background morphology, eventually breaking up the disc into a ring series, in the limit of a strong non-linear back-reaction.

This issue is different -- from the very beginning -- from the standard model of disc dynamics and its well-known open questions, such as the problem of turbulence-enhanced viscosity (for review purposes, you can see \cite{BKL01,B03REVIEW}). The main purpose of this work is to generalize this approach to the time-dependent case, showing that this kind of structures holds locally in space and time: it could be therefore used to address the explanation of local processes and transient phenomena.

The brand new point of view stems from the implementation of an ideal 
MHD scheme where the Ferraro 
Corotation Theorem \cite{F37} holds,
allowing the angular velocity of the disc to be expressed in terms of magnetic flux surfaces only (for a stationary visco-resistive extension of this model see \cite{BM10,BMP11}). 
In fact, the background centrifugal force is balanced by the central 
object gravity -- resulting in the Keplerian rotation of the disc --
as well as by the background magnetic field, since the initial magnetosphere has a 
current-free morphology.
When the back-reaction corrections to the centrifugal 
force and the Lorentz force are expressed in terms 
of the magnetic flux function, their balance results in the formation of toroidal current channels embedded in a radial oscillating magnetic structure.
A remarkable assumption in constructing 
this scenario is the pure rotation of the disc, which is not 
endowed with any poloidal component of the velocity field.

Indeed, in many real astrophysical systems, the disc plasma is not very far from the quasi-ideal behaviour, 
but the following three main points need to be discussed.
(i) Given the mean thermodynamic parameters of the 
plasma in an accreting disc (temperature and number density), 
the kinetic theory uniquely determines finite non-zero 
values of the viscosity and resistivity coefficients. 
Even when the value of such coefficients is rather small, 
the impact of their existence -- i.e. the damping they can induce on the magnetic structure -- 
could not be negligible, due to the 
long lifetime of the accreting systems. \\
(ii) The Shakura model for accretion within a thin disc 
configuration \cite{S73,SS73} relies on very large values 
of the viscosity coefficient, able to balance the angular 
momentum transport responsible for a significant non-zero 
accretion rate. 
We will show how the microstructures framework and 
the Shakura model of accretion are not 
straightly comparable scenarios (e.g. the 
purely rotating disc is unable to directly accrete mass); none the less, 
it is of significant interest to 
understand if the periodic structures
predicted in \cite{C05,CR06} 
still survive in the presence of the visco-resistive effects 
required to account for a turbulent accreting 
plasma. \\
(iii) A non-zero resistivity coefficient is always present in every numerical simulation, at least to reproduce the effect of numerical dissipation due to machine's finite precision \cite{BCFMR11}.
Furthermore, works like \cite{ZFRBM07} include non-ideal terms to explore the coupling of disc and jet physics, aiming at the problem of angular momentum transport.

In this work we include visco-resistive effects to the 
microstructures paradigm and grant the time dependence 
of the magnetic surfaces, needed to preserve the 
consistence of the Generalized Ohm Law.
In fact, as far as we include a resistive contribution, the azimuthal 
component of the Ohm Law acquires a term 
proportional to the toroidal current: this would turn out to be 
unbalanced in a purely rotating configuration, 
unless the non-stationarity of the model ensures 
the presence of a non-zero azimuthal electric field.
In such a non-stationary case, we are able to 
recover a solution consistent with the Corotation Theorem and to provide 
the full consistence of the equilibrium configurations, using rather natural assumptions in the thin disc 
limit and requiring the toroidal component 
of the magnetic field to be negligible
(we adopt a dipole configuration to characterize the 
magnetic properties of the central object).

Focusing on the magnetic back-reaction, but neglecting its 
effect on the mass density distribution, 
we obtain a periodic profile for the disc magnetic field, 
isomorphic to the `crystal profile' studied in \cite{C05}
but for the presence of an exponential damping in time.
Note that this is the most general regime for a non-stationary purely rotating disc, which is not requested to steadily accrete matter; we are dealing with a model where accretion could only exist by means of some intermittent instability (e.g., 3D modes excited along a separatrix \cite{C09}).
The characteristic lifetime of this structure 
depends on the resistivity 
(or equivalently on the viscosity, 
since the Prandtl number is constrained to the unity 
by the consistence of the configuration equations) 
and crucially 
on the spatial scale of the radial 
oscillations. 
The obtained time-scales could aim at accounting for transient 
phenomena: lifetimes can be easily 
fitted in the range of density and temperature typical for stellar accretion discs, 
corresponding to radial scales that preserve the request to deal with 
a local model around a fixed value of the radius (i.e. the length scale of the oscillations 
is much below the disc radial size).
The present analysis is therefore crucial in focusing 
the correct phenomenological scenario to which this new paradigm can be referred to, ruling out this morphology of the magnetic field from the steady-state configuration of a stellar accretion disc.

The paper is organized as follows.
In \Sref{sec:eqns}, we review the fundamental equations of axially symmetric two-dimensional MHD describing an accretion disc embedded in the gravitational and magnetic fields of a central object, taking into account the ones which notably differ from the equilibrium equations.
In \Sref{sec:perturbation}, we develop the perturbation scheme for the considered problem and fix a fiducial radius for the local study, in order to expand the relevant equations up to the first order in the length scale of magnetic perturbation.
In \Sref{sec:solution}, we derive the time-dependent form of separable solutions and recover the compatibility of this form with previously obtained equilibrium configurations.
In \Sref{sec:time}, we give estimations of the lifetime of microstructures, checking the consistence with dynamical requests and establishing that this paradigm can address local transient events rather than steady-state configurations.
Concluding remarks will follow in \Sref{sec:final}.

\section{Relevant equations} \label{sec:eqns}

Neglecting the electron pressure gradient,
the Generalized Ohm Law retains its validity in the same form of stationary MHD:
\begin{equation} \label{GOL}
\vec{E} + \df{1}{c} \<{ \vec{v} \mathbf{\times} \vec{B} } = \resisto \vec{J} \, ,
\end{equation}
obviously meaningful at every different time $t$.
This equation links the main physical quantities (electric and magnetic field $\vec{E}$ and $\vec{B}$, velocity field $\vec{v}$, current density field $\vec{J}$) by means of a microscopical transport coefficient -- the resistivity $\resisto$.
Introducing a scalar electric potential $\Phi$ and a vector magnetic potential $\vec{A}$, we get
\SUBEQ{
\begin{equation}
\vec{E} = - \grad \Phi - \df{1}{c} \partial_{t} \vec{A} \, ,
\end{equation}
\begin{equation}
\vec{B} = \grad \times \vec{A} \, ,
\end{equation}
}
where it is crucial to note that a rotational electric field is generated when the magnetic potential depends on time.
We adopt the magnetic flux surfaces formalism and claim that $\vec{A} = \vec{A} \<{\psi}$, with $\psi\<{\vec{r};t}$ the magnetic flux function.
Since the system is axially symmetric, we can choose a cylindrical coordinate system $\<{r,\phi,z}$, where the $\hat{\mathrm{e}}_z$ axis is the symmetry axis; the magnetic potential and field assume then the following form:
\begin{equation} \label{B_form}
\vec{A} = \df{\psi}{r} \hat{\mathrm{e}}_\phi \; \Longrightarrow \; \vec{B} = \df{\grad \psi}{r} \mathbf{\times} \hat{\mathrm{e}}_\phi \, ,
\end{equation}
and we specify that $\psi\<{\vec{r};t} = \psi\<{r,z^2;t}$ is symmetric under reflection over the equatorial plane $z=0$ -- in agreement with the symmetry of the background hydrostatic equilibrium quantities.
This form leads to strictly poloidal components of the magnetic field $\vec{B}$, for it is possible to show, under the hypotheses of the present work, that if the azimuthal magnetic field is set to zero at the initial time, it will remain zero at every following time.
This could be read \virg{on average}, constraining the turbulence-generated field to preserve a zero mean value if no external component is available; we are decoupling the study from the phenomenon of magnetic field generation due to shearing, assuming that our disc is not able to generate and sustain a coherent and significant azimuthal field.

Since we are dealing with a purely rotating configuration in local approximation (in a narrow annulus around a fixed radius), the azimuthal component of Generalized Ohm Law \eqref{GOL} becomes:
\begin{equation}\label{psi_t}
\DERO{\psi}{t} - \df{c^2 \resisto}{4 \pi} \DELTA{ \psi } = 0 \, ,
\end{equation}
having used the Amp\`{e}re Law to remove the current:
\begin{equation}
\vec{J} = \df{c}{4 \pi} \grad \times \<{ \grad \times \vec{A} } \, .
\end{equation}
The same expression must be obtained from the Induction Equation with the usual derivation,
and this can be done only if the resistivity is nearly a constant, precisely if
\begin{equation} \label{grad_eta}
\abs{ \df{\grad \resisto}{\resisto} } \ll \abs{ \df{ \grad \<{\Delta \psi}}{\Delta \psi} } \, ,
\end{equation}
where $\Delta \<{\cdot}$ is the Laplace operator.
It is worth noting that this work is focused on the inner bulk region of the disc where this condition can hold (i.e. far from the interface between the disc and the surrounding magnetosphere, where the resistivity falls abruptly to zero).
Moreover, the azimuthal component of the Induction Equation:
\begin{equation}
\DERO{B_\phi}{t} = \left[ \grad \times \<{ \vec{v} \times \vec{B} } \right]_\phi - \df{c}{4 \pi} \grad \times \<{ \resisto \grad \times \<{B_\phi \vec{E}rsphi} } \, ,
\end{equation}
in the case of zero azimuthal field, yields to the following constraint:
\begin{equation} \label{I_t}
\grad \omega \times \grad \psi = 0\, ,
\end{equation}
where we have introduced the angular velocity $\omega\<{\vec{r};t}$ such that $\vec{v} = \omega r \vec{E}rsphi$, which turns out to be a flux function $\omega\<{r,z;t} = \omega\<{\psi}$. This result generalizes the Corotation Theorem \cite{F37}, which holds in the visco-resistive framework provided there is no azimuthal component of the central magnetic field.

Finally, we deal with the MHD momentum conservation equation:
\begin{equation} \label{momentum}
\begin{split}
\rho \<{ \partial_{t} \vec{v} + \<{ \vec{v} \mathbf{\cdot} \grad } \vec{v} } = & - \grad p - \rho \grad \chi + \vec{F}_L + \\ & + \visco \left[ \nabla^2 \vec{v} + \df{1}{3} \grad \<{ \grad \mathbf{\cdot} \vec{v} } \right] \, ,
\end{split}
\end{equation}
where $p$ is the thermodynamic pressure, $\chi$ is the gravitational potential of the central object, $\vec{F}_L$ is the Lorentz force, and the viscosity $\visco$ is assumed to be a constant.
Its azimuthal component gives us the evolution law for the angular velocity:
\begin{equation} \label{omega_t}
\DERO{\omega}{t} - \df{\visco}{\rho} \DELTA{ \omega } = 0\, ,
\end{equation}
while the other components retain their stationary forms because of the pure rotation assumption.

\section{Perturbative Scheme} \label{sec:perturbation}

We can now separate the contribution of the background dipole-like magnetic field
from the back-reaction induced by the plasma current, i.e.
\begin{equation}
\psi \<{r,z^2;t} = \psi_0 \<{R_0,z^2} + \psi_1 \<{ r-R_0,z^2;t} ,
\end{equation}
where $ R_0 $ is the fiducial value for the local approximation, centred on $ \abs{r-R_0} \ll R_0 $.
We highlight that the background surface $\psi_0$ is a stationary vacuum solution of Laplace's equation \mbox{(i.e.  $\Delta \psi_0 = 0$)}, so \eref{psi_t} can be written by means of the back-reaction $\psi_1$ only.

Correspondingly, since \eref{I_t} constrains the angular velocity to depend upon the magnetic surfaces (and mainly on the background one), it gains the following local expression:
\begin{equation} \label{ferraro}
\omega \<{ R_0 , r-R_0 , z ; t } \equiv \omega \<{\psi} \simeq \omega\<{\psi_0} + \omega'_0 \psi_1 \, ,
\end{equation}
with
\SUBEQ{
\begin{equation}
\omega \<{ \psi_0 } = \Omega_{\mathrm{K}} \<{R_0} = \sqrt{ \df{G M_{\mathrm{*}}}{R_0^3} } \, ,
\end{equation}
\begin{equation}
\omega'_0 = \left. \DERT{\omega}{\psi}{} \right|_{\psi_0} = \mathrm{const.} \, ,
\end{equation}
}
and we recover the Keplerian profile $\Omega_{\mathrm{K}}\<{r}$ for the background contribution.
Note now that the evolution of the angular velocity is determined by \eref{omega_t}: using the expansion \eqref{ferraro}, it reduces itself exactly to \eref{psi_t} if
\begin{equation} \label{prandtl}
\df{\visco}{\rho} = \df{ c^2 \resisto }{ 4 \pi } \, ,
\end{equation}
which states that the Magnetic Prandtl Number $\mathbb{PR}$ is set to one, a condition also adopted in \cite{BM10} with the purpose of recovering the existence of a solution consistent with the Corotation Theorem.

Furthermore, \eref{prandtl} shows that the height-dependence of $\resisto = \resisto \<{\rho}$ is determined only by the density profile, giving a deeper meaning to \eref{grad_eta}. This can be explained after the introduction of three dimensionless variables defined as follows:
\SUBEQ{
\begin{equation} \label{adim_space}
\bar{r} = k\<{r - R_0}, \quad \bar{z} = \sqrt[4]{3 \beta_0} \df{z}{H_0},
\end{equation}
\begin{equation} \label{adim_time}
\bar{t} = \df{t}{\tau} = \df{k^2 \visco}{\rho_0} t \, ,
\end{equation}
} 
and two dimensionless functions written as
\begin{equation}
Y \<{\bar{r},\bar{z}^2;\bar{t}} = \df{k \psi_1}{\left. {\partial \psi_0}/{\partial r} \right|_0}, \qquad  D\<{\bar{z}^2} = \df{\rho}{\rho_0} \, ,
\end{equation}
where $\rho_0 = \rho\<{\bar{r}=0,\bar{z}=0}$ is the density value on the equatorial plane at the fixed radius, $H_0$ is the half-thickness of the disc, and $k^{-1}$ is the radial scale of the back-reaction flux function $\psi_1$.

The parameter $\beta_0$ used in \eref{adim_space} to scale the height-dependence of magnetic surfaces is the usual plasma $\beta$-parameter, but it takes into account the magnitude of the background field only:
\begin{equation} \label{beta}
\beta_0 = 8 \pi \df{p_0}{B_{z0}^2} = \<{k H_0}^2 \, ,
\end{equation}
where $p_0$ is the background thermodynamic pressure.
We can then write down the components of \eref{grad_eta} to obtain:
\SUBEQ{
\begin{equation}
\df{\partial_{z} \resisto}{\resisto} \simeq \df{1}{H_0} \ll \df{\sqrt[4]{\beta_0}}{H_0} \simeq \df{\partial_{z} \psi_1}{\psi_1}
\end{equation}
\begin{equation} \label{kR0}
\df{\partial_{r} \resisto}{\resisto} \simeq \df{1}{R_0} \ll k \simeq \df{\partial_{r} \psi_1}{\psi_1} \, ,
\end{equation}
}
which are identically satisfied if
\begin{equation} \label{regime}
\beta_0 \gg 1 \, ,
\end{equation}
since \eref{kR0} -- via \eref{beta} -- can be restated as
\begin{equation}
k R_0 \simeq \sqrt{\beta_0} \df{R_0}{H_0} \gg 1 \, ,
\end{equation}
and this is implied by \eref{regime} and the thin disc assumption $H_0 \ll R_0$.
Then it follows that \eref{regime} specifies the regime of validity for our treatment, for it assures the slowly varying behaviour of the resistivity.

In terms of these variables and parameters, \eref{psi_t} becomes
\begin{equation} \label{heateq}
D\<{\bar{z}^2} \DERO{Y}{\bar{t}} - \DELTAE{Y} = 0 \, ,
\end{equation}
where 
\begin{equation}
\DELTAE{\<{\cdot}} \doteq \DER{\<{\cdot}}{\bar{r}}{2} + \dfrac{1}{\sqrt{3\beta_0}} \DER{\<{\cdot}}{\bar{z}}{2}
\end{equation}
is the dimensionless Laplacian.

\section{Damped Solutions} \label{sec:solution}

\begin{figure} \label{fig:crystal}
\centering
\includegraphics[scale=1.3]{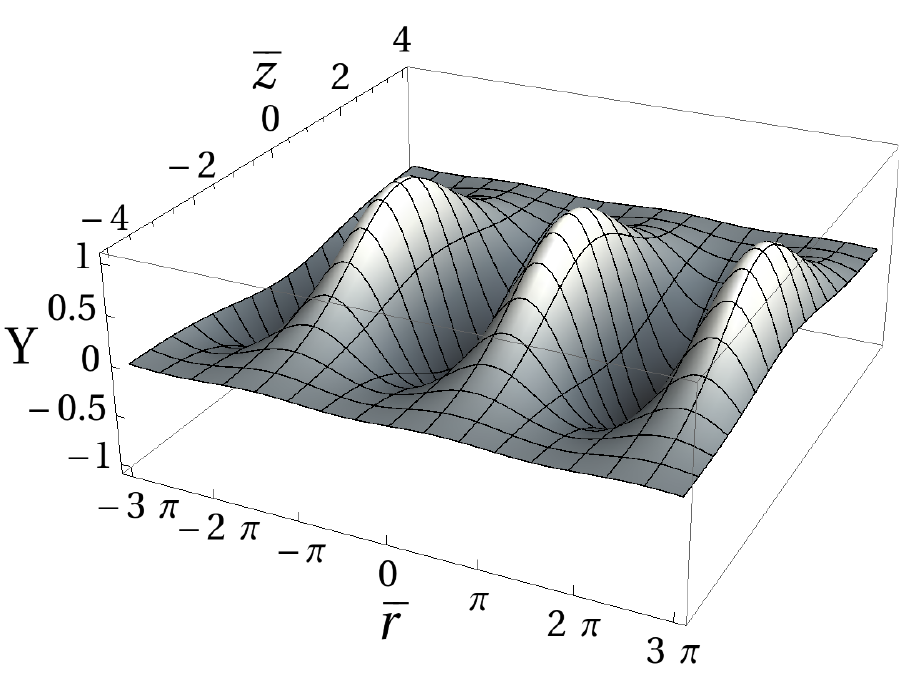}
\caption{Behaviour of the initial ($\bar{t}=0$) back-reaction magnetic surfaces in \eref{solution} versus the dimensionless radial and vertical coordinates $\bar{r}$ and $\bar{z}$, normalized to the maximum equatorial value obtained at the fixed radius. Parameters are chosen to recover the profile of \cite{CR06}, so $\beta_0 = 300/4$, $\Gamma=1$ and $a = a_{\mathrm{min}}$. The rigid oscillatory profile shown is what we referred to as a `magnetic microstructure'.}
\end{figure}

\eref{heateq} is a linear parabolic PDE with an infinite number of solutions: in particular, it admits a separable solution analogue to those found in \cite{CR06}, currently endowed with an explicit time-dependence:
\begin{equation} \label{solution}
Y\<{\bar{r},\bar{z}^2;\bar{t}} = \mathcal{Y} \, F\<{\bar{z}^2} \sin\<{a \, \bar{r}} e^{-\bar{t}} \, ,
\end{equation}
$\mathcal{Y}$ and $a$ being real constants. The height-dependence is fixed by
\begin{equation} \label{F_u2}
\DERT{F\<{\bar{z}^2}}{\bar{z}}{2} - \sqrt{3 \beta_0} \<{a^2 - D\<{\bar{z}^2}} F\<{\bar{z}^2}=0 \, ,
\end{equation}
after the choice of an equation of state, needed -- together with the vertical hydrostatic background equilibrium -- to set an expression for $D\<{\bar{z}^2}$.
If we choose a polytropic form for the gravothermal background, such that:
\SUBEQ{
\begin{equation}
p\<{\bar{z}^2} = p_0 \, D\<{\bar{z}^2}^{1 + {1}/{\Gamma}}
\end{equation}
\begin{equation} \label{D_polytropic}
D\<{\bar{z}^2} = \left[ 1- \<{\df{z}{H_0}}^2 \right]^{\Gamma} \simeq 1 - 	\df{\Gamma}{\sqrt{3\beta_0}} \bar{z}^2 \, ,
\end{equation}
}
where $\Gamma$ is the polytropic index,
then we can find an analytic solution of \eref{F_u2} by means of Generalized Hermite Polynomials
$\mathcal{H}\!\left[ \cdot , \cdot \right]$:
\begin{equation} \label{hermite}
\begin{split}
F\<{\bar{z}^2} = & e^{-\sqrt{\Gamma} \bar{z}^2 / 2} \cdot \\
& \cdot \mathcal{H}\!\!\left[- \df{\sqrt{\Gamma / 3 \beta_0} + a^2 - 1}{2 \sqrt{\Gamma / 3 \beta_0}} , \sqrt{\sqrt{\Gamma} \bar{z}^2} \right] ,
\end{split}
\end{equation}
having omitted a constant factor and adopting the notation $\mathcal{H}\!\left[ \xi , x \right]$ for the Hermite Polynomial of order $\xi$ in the variable $x$.
\begin{figure} \label{fig:profile}
\centering
\includegraphics[scale=.5]{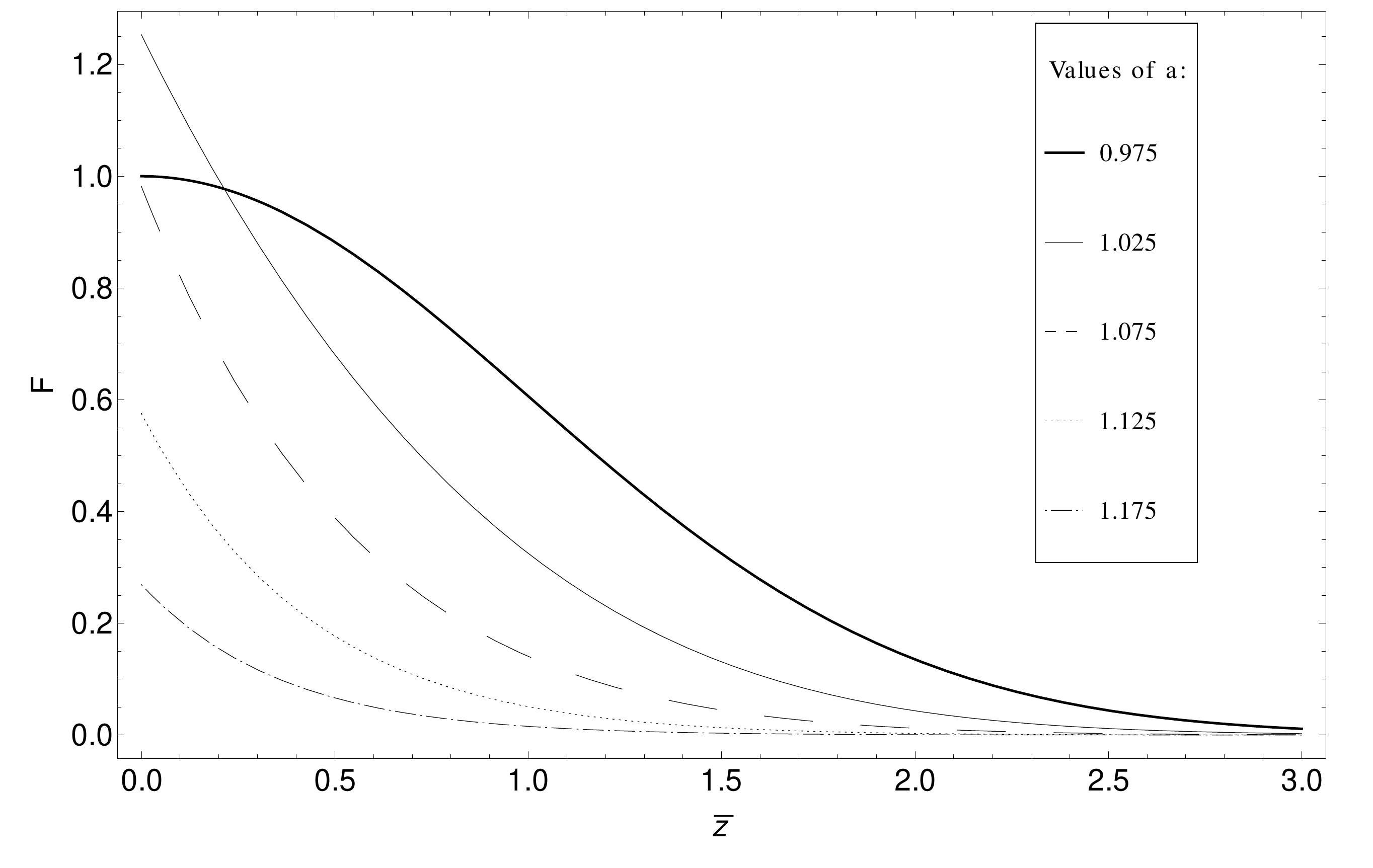}
\caption{Behaviour of the back-reaction magnetic surfaces in \eref{hermite} versus the dimensionless vertical coordinate $\bar{z}$, normalized to the gaussian equatorial value with $a = a_{\mathrm{min}}$.
We fix $\beta_0 = 400/3$ and $\Gamma=1$, and we change $a$ from $ a_{\mathrm{min}} $.
As $a$ increases, the plots are drawn as follows: thick, thin, dashed, dotted, dot-dashed.
It is worth noting that the gaussian profile has the greatest full width at half maximum, while it has not the greatest initial amplitude.}
\end{figure}
These Generalized Polynomials are monotonically decreasing provided their order is negative.
Choosing the radial wave-number as:
\begin{equation}
a^2 = a^2_{\mathrm{min}} = {1-\sqrt{\Gamma / 3 \beta_0}} \, ,
\end{equation}
the solution \eqref{hermite} reproduces the gaussian profile already obtained in \cite{C05}.
This is the lower limit for the range of the parameter values, since the solution becomes non-physical when $a < a_{\mathrm{min}}$; on the contrary, an upper limit doesn't exist, although the amplitude of the structure becomes smaller 
as $a$ increases, and it is eventually negligible slightly over the unity (see \manualref{fig:profile}{1}).
Only a narrow range of values around the unity is therefore suitable to tune the radial wavenumber.
After these considerations, the solution shown in \eref{solution} is plotted in \manualref{fig:crystal}{2} for the isothermal case, as a prototype of a magnetic microstructure.
\subsection*{Matching With Previous Models} \label{sec:matching}
We underline that, under the assumption of pure rotation $\vec{v} = \omega R_0 \vec{E}rsphi$, the only equations explicitly depending on time are those shown in \Sref{sec:eqns}, though there are other interesting equations.

Radial and vertical components of momentum conservation \eqref{momentum} retain their stationary formulation, returning the exact same dimensionless equations offered in \cite{C05,CR06}, now coupled to \eref{heateq}.
This equation introduces the new variable $\bar{t}$, but acts as a closure condition for the system which determines the equilibrium of perturbative pressure $\hat{P}$ and density $\hat{D}$, namely:
\begin{subequations} \label{coppi_eqns}
\begin{equation}
\partial_{\bar{z}^2} \hat{P} + \df{1}{\sqrt{3 \beta_0}} \hat{D} + 2 \DELTAE{Y} \partial_{\bar{z}^2} Y = 0
\end{equation}
\begin{equation}
\begin{split}
\df{1}{2} \partial_{\bar{r}} \hat{P} + \left( D\<{\bar{z}^2} + \df{1}{\beta_0} \hat{D} \right) & Y + \\ + \DELTAE{Y} & \<{ 1+\partial_{\bar{r}} Y} = 0 \, .
\end{split}
\end{equation}
\end{subequations}
In the linear regime $Y \ll 1$, when pressure and density perturbations are assumed to be negligible, system \eqref{coppi_eqns} is fully satisfied solving the equation
\begin{equation}
\DELTAE{Y} = - D\<{\bar{z}^2} Y \, ,
\end{equation}
which is consistent with the separable solution \eqref{solution} and owns the same magnetic structure (although damped in time) developed in \cite{LM10} and \cite{BMP11}, according to the additional approximations considered in those papers.
It is interesting to note that \eref{heateq} could be a closure condition also in the general non-linear regime, but this is forbidden by the Continuity equation:
\begin{equation} \label{continuity}
\DERO{\rho}{t} + \grad \mathbf{\cdot} \<{ \rho \vec{v} } = 0 \, ,
\end{equation}
which shows that the density has to keep its steady-state profile. 
This suggests that a complete model with the disc decomposition in a ring-like structure resembling the one in \cite{CR06} (which needs the density to be time-dependent) should exhibit non-vanishing poloidal velocities.

Another solution can be found in the special `crystal regime' specified by
\begin{equation} \label{crystalregime}
\mathcal{Y} \gtrsim 1 \, , \quad \hat{D} \ll D \, ,
\end{equation}
occurring when the back-reaction field is strong -- it is indeed determined by the constant $\mathcal{Y}$ in \eref{solution} -- whereas density perturbations are negligible. It is worth noting here that background and back-reaction field magnitudes are completely independent, so that \eref{crystalregime} is still consistent with \eref{regime}.
In this case, the system \eqref{coppi_eqns} is reduced to: 
\SUBEQ{
\begin{equation}
D\<{\bar{z}^2} \partial_{\bar{t}} Y - \DELTAE{Y} = 0
\end{equation}
 
\begin{equation}
\partial_{\bar{z}^2} \hat{P} + 2 \DELTAE{Y} \partial_{\bar{z}^2} Y = 0
\end{equation}
 
\begin{equation}
\df{1}{2} \partial_{\bar{r}} \hat{P} + D\<{\bar{z}^2} Y + \DELTAE{Y} \<{ 1+\partial_{\bar{r}} Y} = 0 \, ,
\end{equation}
}
which admits the solution \eqref{solution} and gives a separable form of the pressure too:
\begin{equation} \label{PY^2}
\hat{P} \<{\bar{r},\bar{z}^2;\bar{t}}= \mathcal{Y}^2 \, D\<{\bar{z}^2} F^2\<{\bar{z}^2} \sin^2\<{a \bar{r}} e^{-2\bar{t}} \, ,
\end{equation}
where $D\<{\bar{z}^2}$ and $F\<{\bar{z}^2}$ can be expressed via \eref{D_polytropic} and \eref{hermite}, respectively.
This perturbative pressure increase corresponds to a temperature (or internal energy) increase, eventually modelled by a perturbative $\hat{T} \<{\bar{r},\bar{z}^2;\bar{t}}$, for in this peculiar regime there is no density perturbation.

\section{Estimations of Damping Time} \label{sec:time}

It turns out that every magnetic structure obtained in literature as an equilibrium configuration becomes a dynamical solution after the compatibility with \eref{heateq} has been checked, admitting in this perspective also solutions slightly different from the proposed \eref{solution}. 
At the same time, the damping effect cannot be removed: the unperturbed magnetic configuration dominated by the central field is restored and the back-reaction microstructured field becomes negligible after the time
\begin{equation} \label{life}
\tau = \df{ \rho_0 }{k^2 \visco} \, ,
\end{equation}
defined in \eref{adim_time}.
A direct consequence is that such structures need the plasma to be quasi-ideal; in particular, this shows how this perspective cannot be consistent with the Standard Model \cite{S73,SS73} and its huge effective viscosity, which damps the structures after very little time.

\begin{figure} \label{fig:PR}
\centering
\includegraphics[scale=.45]{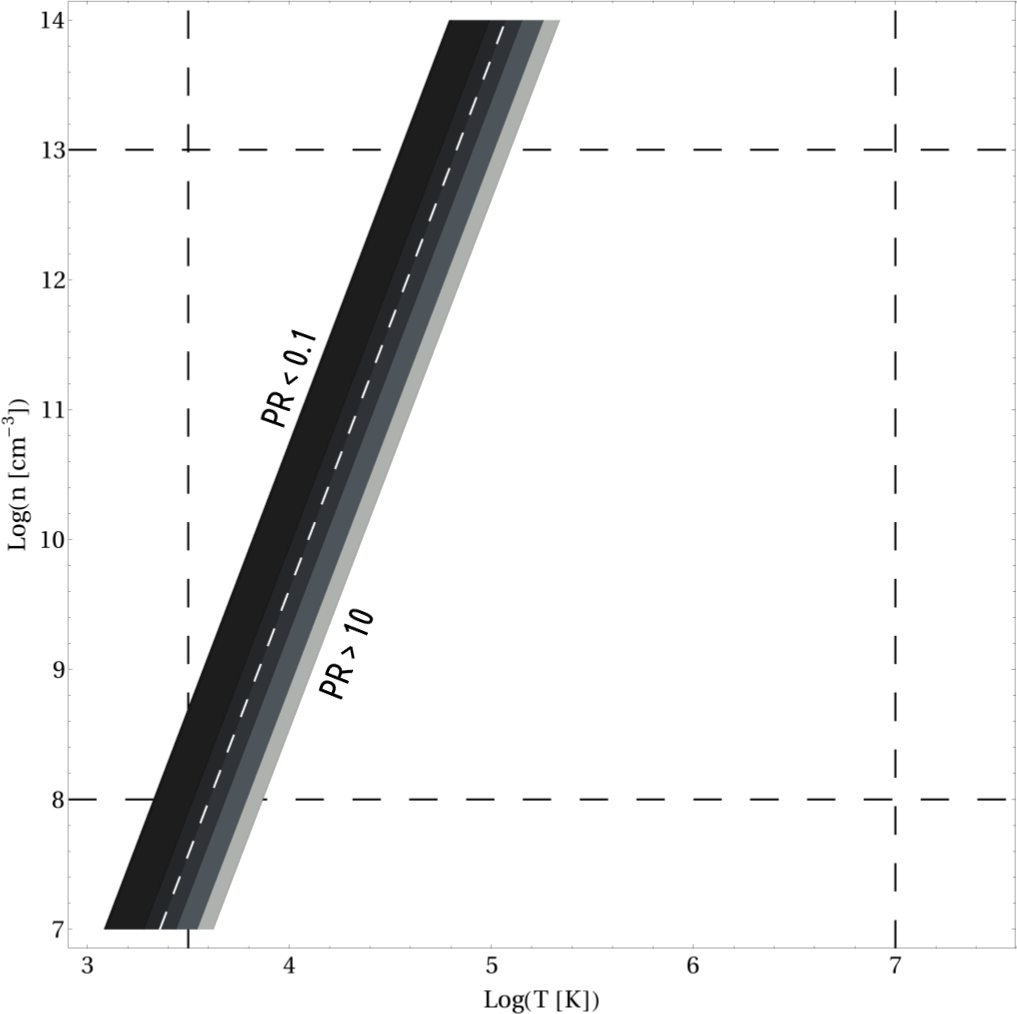}
\caption{
Contour plot of Magnetic Prandtl Number in the log-log plane of temperature and number density via \eref{PRlong}, restricted from values $\mathbb{PR}=0.1$ (darkest stripe) to ${\mathbb{PR}=10}$ (lightest stripe).
The white dashed line is $\mathbb{PR}=1$ and points out the behaviour of the implied relation $n_\mathrm{e}\<{T}$.
The black dashed lines mark the ranges consistent with known accretion discs.
}
\end{figure}

This quasi-ideal lifetime depends mainly on number density $n_\mathrm{e}$ and temperature $T$, and their possible values are interrelated because of \eref{prandtl}.
In what follows we make use of microscopical resistivity:
\begin{equation} \label{microeta}
\resisto = \df{m_{\mathrm{e}} \nu_{\mathrm{ie}}}{n_\mathrm{e}\mathrm{e}^2} \simeq 4 \pi \mathrm{e}^2 \sqrt{\df{m_{\mathrm{e}}}{K_{\scriptscriptstyle \mathrm{B}}^3}} \, \df{\mathrm{Log} \Lambda \<{n_\mathrm{e}, T}}{\left( T \left[\!\, \mathrm{K}\right] \right)^{3/2}} \; \left[\!\, \mathrm{s}\right] \, ,
\end{equation}
and microscopical viscosity:
\begin{equation} \label{microvisco}
\visco = \df{m_{\mathrm{i}} n_\mathrm{e} c_\mathrm{S}^2}{\nu_{\mathrm{ii}}} \simeq \df{\sqrt{m_{\mathrm{i}} K_{\scriptscriptstyle \mathrm{B}}^5}}{4 \pi \mathrm{e}^4} \, \df{\,\<{T \left[\!\, \mathrm{K}\right]}^{5/2}}{\mathrm{Log} \Lambda \<{n_\mathrm{e}, T}} \; \left[\! \df{\, \mathrm{g}}{\, \mathrm{cm} \cdot \, \mathrm{s}} \right] \, ,
\end{equation}
where $m_{\mathrm{e}}$ and $\mathrm{e}$ are electronic mass and charge, $m_{\mathrm{i}}$ is the ionic mass (here protons are considered), $\nu_\mathrm{ie}$ and $\nu_{\mathrm{ii}}$ are the collision frequencies of ions with electrons and ions with ions respectively, $K_{\scriptscriptstyle \mathrm{B}}$ is the Boltzmann constant, and $\mathrm{Log} \Lambda \<{n_\mathrm{e}, T}$ is the Coulomb Logarithm.
By means of these expressions we obtain the quasi-ideal Magnetic Prandtl Number:
\begin{equation} \label{PRlong}
\begin{split}
\mathbb{PR} \doteq & \; \df{4 \pi \visco}{c^2 \rho \resisto} = \\
& \df{K_{\scriptscriptstyle \mathrm{B}}^4}{4 \pi \mathrm{e}^6 c^2 \sqrt{m_{\mathrm{e}} m_{\mathrm{i}}}} \, \df{\left( T \left[\!\, \mathrm{K}\right] \right)^4}{\mathrm{Log} \Lambda \<{n_\mathrm{e}, T} \, n_\mathrm{e}\left[\! \, \mathrm{cm}^{-3}\right]} \, ,
\end{split}
\end{equation}
so that the request $\mathbb{PR}=1$ sets a transcendental relation $n_\mathrm{e} = n_\mathrm{e} \<{T}$, represented as a straight line in \manualref{fig:PR}{3}.
The same Figure shows how the range of possible temperatures and densities shrinks when a real accretion disc is considered.
We deduce that the microstructures -- as described here -- could only be found in correspondence of specified values of the involved physical parameters, i.e.  with densities in $\<{10^8, 10^{13}} \, \mathrm{cm}^{-3}$ and temperatures in $\<{5\cdot10^3, 10^5} \, \mathrm{K}$.

Having assigned the $n_\mathrm{e}\<{T}$ prescription, we are able to check if the microstructures exist beyond the dynamical time-scale $\Omega_{\mathrm{K}}^{-1}$, which is the time needed for the vertical hydrostatic equilibrium to be established: since we assumed it is preserved (vertical background equilibrium equation), the microstructures have to exceed it for the model to be consistent. 
The \eref{life} is recast in this terms as:
\begin{equation} \label{tauomega}
\df{\tau}{\Omega_{\mathrm{K}}^{-1}} = \<{\df{\lambda}{2 \pi c_\mathrm{S} \<{T}}}^2 \nu_{\mathrm{ii}}\<{T} \Omega_{\mathrm{K}}\<{M_\mathrm{*},R_0} \, ,
\end{equation}
where $\lambda \doteq 2 \pi k^{-1}$ is the back-reaction length scale.

\begin{figure} \label{fig:tauomega}
\centering
\includegraphics[scale=.5]{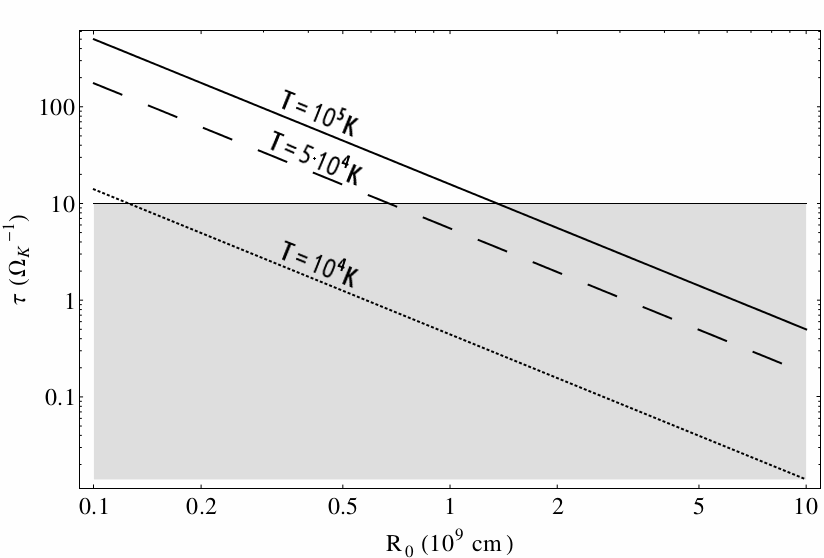}
\caption{
Plot of microstructures lifetime versus fiducial radius $R_0$ via \eref{tauomega}, with central mass $M_\mathrm{*} = 1 M_\mathrm{\odot}$.
Microstructures length scale adopted here is $\lambda = 5 \cdot 10^3 \, \mathrm{cm}$, following the estimation in \cite{MB11}.
Time is in units of dynamical time, so the thin horizontal line marking the shaded region acts as the threshold for $\tau \gg \Omega_{\mathrm{K}}^{-1}$.
The $\tau\<{R_0}$ lines are drawn for $T=10^4 \, \mathrm{K}$ (dotted line), $5\cdot10^4 \, \mathrm{K}$ (dashed) and $10^5 \, \mathrm{K}$ (solid).
}
\end{figure}

Seeking for the consistence check, we make use of an estimation provided for an analogous global structure \cite{MB11}, which is of the order of $\<{10^3,10^4} \, \mathrm{cm}$.
In \manualref{fig:tauomega}{4} we then represent the lifetime at a fixed $\lambda$ as a function of the radial parameter $R_0$, for different values of the temperature $T$; we deduce that this model has not to be applied to cold discs with a mean temperature below $10^4 \, \mathrm{K}$.
Magnetic microstructures can appear in the inner regions of discs with $T$ in $\<{10^4,10^5} \, \mathrm{K}$.
Looking at the details of \manualref{fig:tauomega}{4}, we note that at the lowest temperatures the lifetime lies in the shaded region, where microstructures vanish before vertical equilibrium is established or their rise is competitive with Magneto-Rotational Instability ($\tau \lesssim \Omega_{\mathrm{K}}^{-1}$) \cite{BH91}, so there is no formation of periodic magnetic flux surfaces.
Raising the temperature, the lifetime eventually crosses the threshold such that $\tau \gg \Omega_{\mathrm{K}}^{-1}$, starting from the lower values of radius: at $T=10^5 \, \mathrm{K}$ the structure is however confined within radii less than $10^9 \, \mathrm{cm}$, and the highest temperatures are forbidden by the constraint $\mathbb{PR}=1$.

We have also to remember the request of \eref{regime} on the $\beta_0$ parameter, which in this narrow range of temperatures would imply a limitation on the background magnetic field magnitude.
This turns out to be easily feasible since it only asks the vertical magnetic field to not exceed $10^{12} \, \mathrm{G}$, ruling out only the strongest of the magnetars; even highly magnetized neutron stars reach this value only at their surface, with little or no influence on the disc.

\begin{figure} \label{fig:tau}
\centering
\includegraphics[scale=.5]{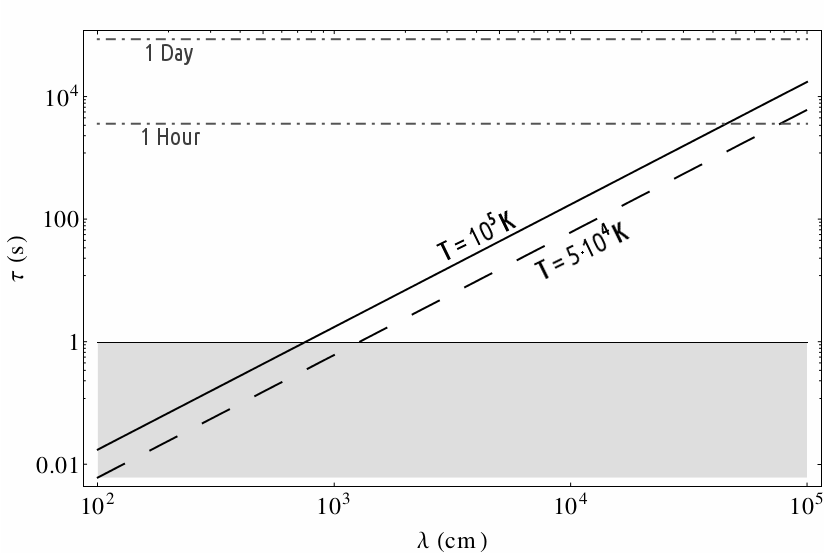}
\caption{Plot of microstructures lifetime versus their length scale via \eref{real_lifetime}.
The horizontal thin line marking the value of dynamical time $\Omega_{\mathrm{K}}^{-1}\<{M_\mathrm{*},R_0}$ is drawn after fixing the values of central mass and disc scale radius to $M_\mathrm{*} = 1 M_\mathrm{\odot}$ and $R_0 = 5\cdot10^8 \, \mathrm{cm}$.
The $\tau\<{\lambda}$ lines are drawn for $T=5\cdot10^4 \, \mathrm{K}$ (dashed) and $10^5 \, \mathrm{K}$ (solid).
The dot-dashed horizontal lines mark the value of one hour and one day respectively, in order to show the lifetimes order of magnitude, which is mainly from minutes to hours and never exceeds the day.
}
\end{figure}

Microstructures are then characterized locally in space (inner disc region) and time (finite lifetime).
In a more quantitative way, fixing the fiducial radius and cutting the lowest values of temperature, lifetime can be estimated in seconds as a function of the microstructures length scale (note that a fixed radius implies a fixed dynamical time-scale).
It simply stems from \eref{life} again by means of \eref{microvisco}:
\begin{equation} \label{real_lifetime}
\tau
= \df{\mathrm{e}^4}{\pi} \sqrt{\df{m_{\mathrm{i}}}{K_{\scriptscriptstyle \mathrm{B}}^5}} \left( \lambda \left[\!\, \mathrm{cm} \right] \right)^2  \df{\mathrm{Log} \Lambda \<{n_\mathrm{e}, T} \, n_\mathrm{e}\left[\! \, \mathrm{cm}^{-3}\right]}{\left( T \left[\!\, \mathrm{K}\right] \right)^{5/2}} \, ,
\end{equation}
giving at the same time a lower bound for the scale of the spatial fluctuations of back-reaction field, as it is represented in \manualref{fig:tau}{5}.
Existing structures have size of at least tens of meters, outlive the dynamical time and vanish over minutes.
Raising the temperature, smaller structures emerge with the same lifetime, while the biggest scales reach lifetimes of the order of hours.
It is worth noting that such scales are tunable, in the sense that the model cannot fix neither $\lambda$ nor $\tau$ because of the assumption of locality.
Transient events whose duration lies in the range of minutes-hours has been observed in accretion discs, so we infer they can be related to the formation of microstructures.

\subsection*{Relaxing The Prandtl Constraint}
It is possible to preserve the $\mathbb{PR}=1$ condition and avoid the ranges-shrinking at the same time.
Let us adopt an effective resistivity, enhanced because of the far bigger value of viscosity, and parametrized in the form:
\begin{equation} \label{effective_eta}
\resisto^{\mathrm{eff}} = \df{4 \pi}{c^2 \rho} \visco \, ,
\end{equation}
preserving in this way the validity of \eref{prandtl}, regardless of the density-temperature relation determined in the quasi-ideal case.
With this assumption, the accessible values of the physical parameters are broadened up to the full ranges consistent with observations.
It is then possible to get the same lifetimes in a way more extended range of physical states, and even reach time-scales of the order of days for peculiar dense and cold discs.

It is useful to say that the Prandtl condition could also be avoided if some other model-feature is included: e.g. a radial component of the velocity field would keep the same perspective on finite lifetime, but allowing $\mathbb{PR} \neq 1$ \cite{MC12}.

\section{Final Remarks} \label{sec:final}

We discussed how the equilibrium configurations proposed by \cite{C05} and the following works have to be generalized by means of non-ideal terms (an arbitrarily small but non-zero resistivity is required by kinetic theory, and its effects are not negligible), and how these equilibria evolve in time.
The peculiar morphology of magnetic microstructures is preserved, but it has been shown how the dissipative coefficients forbid its maintaining: in particular, the macroscopical viscosity of Shakura Standard Disc destroys it on a time-scale so much short that the structure formation itself is forbidden.
The microscopical plasma viscosity gives instead the chance to outlive the dynamical threshold: quasi-ideal microstructures vanish after minutes and eventually hours.
This time-scale is consistent with the description of transient events, once the model is refined to gain a perspective in the emissions possibly related to localized structures of magnetic field and induced current.

This result rules out the stationary morphology in stellar accretion disc (with a central neutron star or a solar mass black hole) adopted until now, because of the finite lifetime found also in the quasi-ideal case.
Realistic \virg{crystal structures} can exhibit a length scale of at least $10^3 \, \mathrm{cm}$, confined by the request to exceed the dynamical time (lower bound).
It is clear at this point that the aimed transient events have a local nature in both space and time -- spatially confined to microscales and temporally limited by the structures mean lifetime.

It has also been shown that only temperatures between $10^4$ and $10^5 \, \mathrm{K}$ allow the formation of quasi-ideal structures, and this formation is easier in the inner region of discs, up to a radial extent of the order of $10^9 \, \mathrm{cm}$.
For central magnetic fields stronger than $10^{12} \, \mathrm{G}$ structures are forbidden, but only magnetars could reach such an order of magnitude, while the range of temperatures and sizes is consistent with observed or estimated parameters in binary accretion discs with a magnetized central object (white dwarf, neutron star or low mass black hole), also with a meaningful magnitude of the magnetic field far from the object surface.

\paragraph*{Acknoledgments.}
This work has been partially developed in the framework
of the CGW Collaboration (www.cgwcollaboration.it).


%

\end{document}